\documentstyle[aps,12pt]{revtex}
\begin{document}
\author{A. Tartaglia}
\address{Dip. Fisica, Politecnico, Torino, Italy\\
E-mail tartaglia@polito.it}
\title{Lengths on rotating platforms}
\maketitle

\begin{abstract}
The paper treats the issue of the length of a rotating circumference as seen
from on board the moving disk and from an inertial reference frame. It is
shown that, properly defining a measuring process, the result is in both
cases $2\pi R$ thus dissolving the Ehrenfest paradox. The same holds good
when considering that, for the rotating observer, the perceived radius
coincides with the curvature radius of a space-time helix and a complete
round trip corresponds to an angle which differs from the one seen by the
inertial observer. The apparent contradiction with the Lorentz contraction
is discussed.
\end{abstract}


\section{Introduction}

There is a problem in special relativity which runs across all the history
of the development of this fundamental theory up to our days: it concerns
length and time measurements on a simple uniformly rotating platform.

A special relativistic approach, moving from the Lorentz contraction along
the direction of motion and the possibility of describing events happening
in accelerated frames in the instantaneously comoving inertial frame, lead
Einstein himself \cite{einstein} and many others \cite{altri} to the
conclusion that the ratio between the length of a rotating circumference and
its radius, either as seen by an inertial observer at rest with respect to
the rotation axis or by a co-rotating observer, cannot be $2\pi .$ A problem
then arose because this result would imply that the space of a rotating disk
is curved, but an observer on board the disk would not perceive any
curvature since in his reference frame both circumference and radius would
have remained the same: this is known as the Ehrenfest paradox\cite
{ehrenfest}.

In the literature the issue continued to be discussed now and then, mostly
assuming as a fact the Lorentz contraction of the circumference of the
rotating disk either as seen from the inertial reference frame or as seen on
board the disk itself. Moving from this start inevitably some kind of
paradox arises; the last one, pointed out by Selleri\cite{selleri},
challenges the very foundation of special relativity. Selleri's case is
solved when considering a fully four-dimensional scenario\cite{rt}, but the
question about what the length of the circumference in the various reference
frames actually is, remains open.

People who move from a ''hard '' interpretation of the Lorentz contraction
are lead to discuss the space curvature of the disk and sometimes to
consider ''relativistic stresses'' and physical deformations or to find
dynamical explanations\cite{phipps}, which, far from solving the problem,
introduce new elements of contradiction and ambiguity.

Here the problem is being rediscussed starting from what a direct length
measurement is and from what one actually measures when looking at a moving
target. Any real measurement, in particular any measurement concerning the
geometry of the disk, is an operation on an extended region of space,
whereas in accelerated reference frames special relativistic formulae are
merely local: the results obtained by generalizing them globally are most
often flawed by contradictions and paradoxes.

The author's view is that the inertial observer can only {\it measure }%
lengths and time intervals using rods and clocks of his own reference frame.
He simply {\it infers} results relating to the rotating platform on the
basis of some mental operation performed starting from the actual
measurements. Starting with this approach and adopting a specific
operational procedure to measure times and lengths we shall see that the
paradox disappears.

\section{Measuring the circumference and the radius}

Consider a non rotating platform in an inertial reference frame. Suppose the
platform is lodged within a coaxial circular hollow with a negligibly
greater radius, dug in a plane plate. An observer places a set of rulers
tail to head both along the rim of the platform and the rim of the hole:
their number is a practical measure of the length of the circumference and,
considering the circumstances, the result is essentially the same both as
seen from the plate and as seen from the platform (be it $L_{0}$). Rulers
are similarly laid down along one radius of the platform: the ratio between
the number along the rim and the number along the radius (this we shall call 
$R_{0}$) is close to $2\pi $ (it would be exactly $2\pi $ if the rulers were
infinitesimal in length). To mark the starting point along both
circumferences (on the disk and on the fixed base) a couple of poles are put
up vertically.

\begin{figure}[tbp]
\caption{Fig.1 Equal meter sticks are laid along two almost identical
circumferences, one of which is the rim of a disk to be successively set in
motion at a constant angular speed. A couple of vertical stick mark the site
of the origin of both circumferences.}
\end{figure}

\subsection{The inertial observer}

Now the observer boards the platform and a friend of his remains on the
static floor, then the platform is set into steady rotation.

Let us start considering what the inertial friend sees and measures. First
he notices that the tiny gap between the rim of the turntable and its fixed
circular housing is always the same as before and consequently he concludes
that the diameter of the rotating table is unchanged: $2R_{0}$. Then our
fellow sets out to measure the length of the rotating circumference or, at
least, to compare the rotating circumference, globally taken, with a
circumference at rest with respect to him, whose length he knows. This
procedure by comparison is, after all, the standard one that is used for
direct measurements. Again the fact that the radial distance between moving
and fixed circumference is unmodified makes him conclude that the sought for
length is the same as the one of the fixed circle and consequently the same
as before: $L_{0}$ . The ratio between circumference and radius length is
still roughly $2\pi $. On the inertial observer's opinion nothing happened
to the geometry of the turntable because of the rotation (we are going to
discuss this point later on).

Finally our experimenter takes a good clock and stands by the fixed fiducial
pole to measure the rotation time of the moving platform. Actually he
registers the time interval between two successive passages of the rotating
pole in the vicinity of the fixed one and finds, let's say, $T_{0}$. The
conclusion is that the rotation speed of the platform is $\omega _{0}=\frac{%
2\pi }{T_{0}}$. The peripheral velocity is of course $v_{0}=\omega _{0}R_{0}=%
\frac{2\pi R_{0}}{T_{0}}$ fully consistent with a circumference length equal
to $2\pi R_{0}$. Indeed, since the pole on the disk is practically moving
along the fiducial (static) circumference, the length it came across when
reaching back the fixed starting point is most naturally $2\pi R_{0}$.

Actually in the literature one finds two different attitudes toward the
value the inertial observer should attribute to the moving circumference.

One approach (see for instance Dieks\cite{dieks} and the many earlier
references quoted in Anderson, Vetharaniam and Stedman\cite{anderson}) is
based on the consideration that any small portion of the circumference, seen
by the inertial observer, is Lorentz contracted. Consequently, when adding
up all of these segments, the whole moving circumference should appear to
the said observer as being Lorentz contracted too: $2\pi R_{0}\sqrt{1-\frac{%
\omega ^{2}R_{0}^{2}}{c^{2}}}$. This result contradicts what our observer
actually measures by his meter rods and would lead, if it were accepted, to
the peripheral speed 
\begin{equation}
v=\frac{2\pi R_{0}}{T_{0}}\sqrt{1-\frac{\omega ^{2}R_{0}^{2}}{c^{2}}}
\label{errata}
\end{equation}

However, by definition, it is also $v=\omega R_{0}$. Solving then (\ref
{errata}) for $\omega $ one obtains 
\begin{equation}
\omega =\frac{2\pi }{\sqrt{T_{0}^{2}+\frac{4\pi ^{2}R_{0}^{2}}{c^{2}}}}
\label{omega}
\end{equation}

Expression (\ref{omega}) shows an interesting behaviour with $T_{0}$ since
when it tends to zero $\omega $ goes to $c/R_{0}$ which is the maximum
allowable angular velocity. Unfortunately however there is also a dependence
of $\omega $ on $R,$ which should show up throughout the disk. The inertial
observer easily verifies that there is no differential rotation and, as
well, time for him is a coordinated (global) time. Now, since what is really
measured by direct comparison with a (circular) meter is the global length
and since time is read on perfectly synchronizable clocks giving the same
result at any point in space, the easiest conclusion is that the
circumference length has not changed.

The second approach, which is the most widely spread (see for instance \cite
{einstein}, \cite{arzelies}, \cite{anandan}... ), assuming that of course
the Lorentz contraction affects the meter sticks on the platform, leads to
the conclusion that the measure of the rotating circumference in the
inertial reference frame should be greater than before (precisely because
the circumference has not changed whilst the meter stick has shortened).
Then, since the radius should remain the same, the ratio between
circumference and radius should be greater than $2\pi $. This conclusion is
incorrect because it is obtained comparing an unmodified circumference
length with rods which are not at rest with respect to the inertial
reference frame. On the other hand on board the platform both the rim and
the rods are at rest and no contraction occurs. In general the inertial
observer cannot use the measuring rods of another (moving) reference frame
to measure lengths in his own reference frame nor attribute length
measurements in another reference frame using the unit rods of that frame 
{\it as seen in his own}.

\subsection{The rotating observer}

Let us consider the view point of the observer on the rotating platform. To
measure the length of the rim of the turntable he counts the number of
rulers along it: this could not change when the disk was set in motion,
since the rulers are laid tail to head and none of them contract or expand
as seen by an observer at rest with respect to them. Result: in our friend's
opinion the length of the circumference is, as before, $L_{0}$.

The same conclusion, for the same reasons, is attained regarding the radius,
which turns out still to be $R_{0}$. The ratio $L_{0}/R_{0}$ is of course $%
2\pi $: in the rotating observer's view nothing happened to the geometry of
the platform as a consequence of the rotation.

At this point the moving observer decides to determine the rotation speed of
his platform. For this he stands near the moving upright pole bearing a
clock identical to his inertial friend's one. Our experimenter measures the
time interval between two successive passages of the pole on the platform by
the fixed inertial one: the result is $T\left( R\right) $ which now differs
from the measure obtained in the inertial reference frame: $T\left( R\right)
<T_{0}$. For the passenger on the rim of the turntable the rotation speed is 
$\omega \left( R\right) =2\pi /T\left( R\right) >\omega _{0}$.

The measurement of the rotation period may be repeated, with the same
technique, at different radii on the platform. For $0<r<R$ the result is $%
T\left( R\right) <T\left( r\right) <T_{0}$, then also $\omega \left(
R\right) >\omega \left( r\right) >\omega _{0}$. Only at the center of the
disk it is $T\left( 0\right) =T_{0}$ and $\omega \left( 0\right) =\omega
_{0} $. Our rotating experimenter however easily verifies that the relative
angular separation between the points used to determine, at different radii,
the angular speed is not varying with time: the platform under his feet is
and remains solid. He then concludes that there is no differential rotation
on the disk, but instead identical clocks tick differently at different
radii: they run slower on the rim than at the center, where the time
intervals coincide with those in the inertial frame.

Empirically or, to say better, experimentally (though with thought
experiments) the only thing on the turntable which is affected by the
rotation is time. This picture in any case is in principle falsifiable by
experiment.

\section{Full space-time view}

With the above facts established, we may turn to the theoretical
interpretation adopting a special relativistic attitude. This means looking
at the situation from a four-dimensional viewpoint; actually $2+1$
dimensions is enough: the $z$, or axial coordinate of the disk, is
irrelevant.

Now we know that clocks measure spacetime ''lengths'' along the timelike
worldline of the clock itself. This means that the time interval shown by an
inertial clock at rest with respect to the axis of the disk corresponds to a
straight vertical line in a typical space-time diagram (see fig.2). On the
other hand the worldline of a clock on the rim of the rotating turntable is
a helix (fig.2) whose line element is \cite{rt} 
\[
d\tau =\frac{d\theta }{\omega _{0}}\sqrt{1-\beta ^{2}}=dt\sqrt{1-\beta ^{2}} 
\]

Here $\theta $ is the rotation angle as measured in the inertial frame, $%
\beta =\omega _{0}R/c$ and $t$ is the inertial time. The total round trip
time is found integrating over $2\pi $, which amounts to coming back to the
(quasi) intersection with the vertical worldline of the static clock: 
\begin{equation}
T\left( R\right) =\frac{2\pi }{\omega _{0}}\sqrt{1-\beta ^{2}}=T_{0}\sqrt{1-%
\frac{\omega _{0}^{2}R^{2}}{c^{2}}}  \label{tdR}
\end{equation}

\begin{figure}[tbp]
\caption{Fig.2 $2+1$ dimensional view of the rotating disk and its static
counterpart. The vertical line is the world line of an observer on the rim
of the fixed circumference; the reading of his clock is proportional to the
length of the vertical segment. The helix is the world line of an observer
on the rim of the rotating platform; his clock measures a time proportional
to the length of the helix. The broken line is the locus of events
''simultaneous'' to the origin.}
\end{figure}

Of course if the moving clock is placed at a different radius $r<R$ it is 
\begin{equation}
T_{0}>T\left( r\right) =T_{0}\sqrt{1-\frac{\omega _{0}^{2}r^{2}}{c^{2}}}%
>T\left( R\right)  \label{tdr}
\end{equation}

This is perfectly consistent with special relativistic formulae (time
dilation) and is due to the fact that the flat space-time is Minkowskian and
the helixes describing the points of the rotating platform are steeper close
to the center than at the periphery. The (apparent) local angular speed $%
\omega $ represents geometrically the instantaneous rotation speed along a
space-like helix, orthogonal to the world line of the observer, and around
its curvature center 
\[
\omega \left( r\right) =\frac{\omega _0}{\sqrt{1-\frac{\omega _0^2r^2}{c^2}}}
\]

since, in complex space-time, it is 
\[
\gamma =\frac 1{\sqrt{1-\beta ^2}}=\cosh \chi 
\]
where $\chi $ is the inclination angle of the helix with respect to a space
hyperplane orthogonal with respect to the inertial time axis.

As far as the actual length measurements which have been performed, they do
not correspond to determining the length of spacelike lines in $2+1$
dimensions. Rather they are counting operations. To clarify this point let
us consider the world lines representing the midpoint of each ruler: whoever
is {\it measuring }the distance between a point on the rim and the center of
the disk is in fact counting the number of such worldlines he crosses on his
way from the periphery to the center. There is indeed a given number of
coaxial cylinders between center and rim, on which the helixes corresponding
to the midpoints of the rulers are wrapped: this number is invariant with
respect to the angular velocity and does not depend on the path one follows
i.e. on the kind of synchronization. Of course the first move, when the disk
was at rest in the inertial frame, was to measure the length of the radius
as a segment of a space geodetic.

Similar considerations hold true when moving along the rim to determine its
length, though this time the line to be measured was not and is not a
geodetic, but was and remains a constant curvature line.

This measuring procedure, which seems quite natural for experimenters living
on the platform, does not produce the length of the locus of events
simultaneous, according to the standard synchronization criterion, with the
one assumed as the departure point. That locus is a portion of a helix and
its length is\cite{rt} 
\[
l_{o}=\frac{2\pi R}{\sqrt{1-\beta ^{2}}} 
\]

This quantity however may only be reckoned in a full space-time perspective;
it is not directly accessible to any observer. Nonetheless for the physicist
on the platform that could be thought to be the rim of the disk, the more
when it is a constant curvature line. From elementary geometry we see that
the curvature radius of a helix in euclidean space ($\alpha $ is the tilt
angle of the instantaneous rotation axis along the helix) is

\[
\rho _{e}=\frac{R}{\cos ^{2}\alpha } 
\]

which, in our case (Minkowski space-time), is translated into: 
\begin{equation}
\rho =\frac{R_{0}}{\cosh ^{2}\chi }=R_{0}\left( 1-\beta ^{2}\right)
\label{curvatura}
\end{equation}

Now our physicist perceives our threedimensional constant curvature
space-time helix as a constant curvature bidimensional line, i. e. a
circumference, whose radius is the radius of the osculating circle\footnote{%
To say better we can speak of an osculating circumference only in complex
space-time and for a spacelike world line. In real space-time and for
timelike world lines the osculating curve is actually a hyperbola.} in a
point of the helix, $\rho $. When describing his round trip survey of the
rim he does the same as if he moved along the osculating circle, though the
latter, during the tour, is precessing around the axis of the cylinder on
which all helixes are wrapped. Of course in our friend's view the round trip
corresponds to a $2\pi $ angle; to work out the relation between his angles, 
$\theta ^{*}$, and the inertial rotations $\theta $ one must consider first
the elementary length of a portion of the simultaneity helix, $%
dl_{0}=R_{0}d\theta /\sqrt{1-\beta ^{2}}$\cite{rt}, then the fact that a
complete round trip of the rotating experimenter corresponds to the inertial
rotation $2\pi /\left( 1-\beta ^{2}\right) $. Compounding together these two
facts leads to: 
\begin{equation}
d\theta ^{*}=\frac{d\theta }{\left( 1-\beta ^{2}\right) ^{3/2}}
\label{angoli}
\end{equation}

The elementary arc length along the rim is for the moving observer 
\begin{equation}
dL=\rho d\theta ^{*}  \label{uno}
\end{equation}

and, since $\rho $ is a constant, it is seen as a circumference element; the
fact that the numerical value attributed to $\rho $ is still $R_{0}$ does
the rest. Converting (\ref{uno}) into inertial frame variables via (\ref
{curvatura}) and (\ref{angoli}) gives 
\begin{equation}
dL=\frac{R_{0}d\theta }{\sqrt{1-\beta ^{2}}}  \label{due}
\end{equation}

which means that the length of the osculating circle is, for the inertial
observer, clearly different from $2\pi R_{0}$, but to him it is also
manifest that this is not the length of the circumference, that he measures
to always be precisely $2\pi R_{0}$.

\section{Discussion}

The long and old controversy regarding the length of a rotating
circumference appears to have here its simplest solution: nothing happens.
What about the Lorentz contraction and its effects?

The point is that all measurements described in this paper and naturally
corresponding to what anyone would actually do, are performed at one point
in space and do not need any particular choice regarding synchronization of
clocks in different places, since the clock used is always one and the same.
The Lorentz contraction is a manifestation of the relativity of simultaneity
in different inertial frames: when no synchronization is needed, no
contraction appears.

Suppose a finite number of inertial observers are placed at equal intervals
along the hollow within which the rotating platform is lodged. They should
synchronize their clocks, then they should read time when a given ruler on
the turntable (let us say the one having the same color as the fixed one
before their feet) passes in front of them: the rulers too are equally
spaced along the rotating contour. When considering the results, they see
that,according to the standard theory, the registered events appear not to
be simultaneous, rather there is an increasing delay in the sense of
rotation and, extrapolating the reading of the last of them toward the first
one, they find he too should be delayed with respect to himself by a time $%
t_{0}=\frac{2\pi \omega _{0}R^{2}}{c^{2}-\omega _{0}^{2}R^{2}}$!

Actually in his reference frame an observer directly measures lengths by
comparison with a gauge; doing so he compares the stationary circumference
with the rotating one and finds they are equal. Then the observer can check
the positions in space-time of some events which, in his frame, should be
simultaneous: on the basis of this simultaneity he infers the fact that the
moving unitary rods, as he sees them, are contracted. Now multiplying the
contracted unit lengths for the number of rods he is led to contradiction
with the direct measurement of the circumference. To solve the contradiction
he must conclude that there is something to be clarified concerning time
measurements and the relativity of simultaneity.

In fact when the same simultaneity criterion is used on the disk as in the
inertial reference frame the studied events are distributed along a
spacelike helix whose pitch is precisely $t_{0}$ \cite{rt}: the clocks of
the inertial observers are measuring the ''lengths'' of vertical spacetime
segments intercepted between the horizontal inertial simultaneity hyperplane
and the points of the said helix. The relevant portion of the helix is the
one going from the start back to the world line of the rotating observer,
winding, in order to reach the intersection, for more than $2\pi $. When
this wrapped-for-more-than-$2\pi $ helix is projected along the world line
of the rotating observer to produce the result seen by the inertial
observer, one actually has a contraction that leads back precisely to a $%
2\pi R$ length.

Even in translational motion the Lorentz contraction can be interpreted this
way . Now the need for a full space-time view is made evident by the fact
that the considered trajectory is closed, which actually means wrapped about
an axis.

We can summarize things still another way. The physicist on the disk always
measures the radius as the curvature radius of the constant curvature line
he sees as the (circular) rim of the disk. When the disk is at rest this
radius coincides with $R_0$, the one of the cylindrical world tube
representing the contour of the platform. When the disk rotates the
curvature radius changes according to (\ref{curvatura}), but the same change
intervenes to the meter sticks the experimenter uses, then the final
numerical result remains the same as in the static case.

As for the static inertial observer, he lives and performs measurements on a
horizontal hyperplane and necessarily finds the length of the circumference
given by the right intersection of his hyperplane with the world tube of the
rim of the (rotating or nonrotating) disk.

Only a fully four dimensional observer (able to perceive globally spacelike
as well as timelike intervals) equipped with invariant fourdimensional meter
sticks will see the rim as an open portion of a helix and find that its
length is of course different from $2\pi R_0$. On the other hand he will
perfectly be able to relate his findings to the previous results without any
contradiction.

Up to this moment we considered two points of view: one for the accelerated
observer on the platform, another for the inertial experimenter at rest with
respect to the rotation axis. There is however a third viewpoint usually
considered either explicitly or implicitly: it is the one of an inertial
observer instantaneously comoving with the man on the platform. To this
observer the non rotating disk would no longer appear as circular, rather it
would be elliptical in shape. If the platform is rotating the situation does
not change: the contour is elliptical and being practically coincident with
the ellipse of the rim of the non rotating lodgement, its length, as well as
all other geometrical parameters are the same. Once again the rotation does
not affect the geometry, simply the simultaneity hyperplanes of the new
observer are skew with respect to the axis of the cylinder on which the
worldlines of points on the rim of the turntable wrap, their sections of the
cylinder are consequently elliptical.

These pictures, though the initial argument was independent of a specific
convention for synchronization, are based on the standard one, the one which
Minkowski\cite{minkowski} himself considered as ''necessary'' 90 years ago.

More formally the procedure that leads traditionally to the idea of a
shortened circumference is based on a twofold passage: first, pass from the
(unique) accelerated (rotating) reference frame to one locally comoving
inertial frame; second, pass from the comoving inertial frame to the
inertial frame of the static observer. The point which seems to be the
origin of the interpretation difficulties is in the fact that the two
extreme reference frames are unique, whereas the intermediate is an infinity
of different comoving inertial frames. What our treatment evidences is the
questionability of the integration procedure based on this infinity of
different frames, or at least of the possibility to interpret the result of
this integration as the length of the rim of the rotating disk. The result
of a calculation, though correct, cannot be considered equivalent to an
actual measurement unless all phases of the calculation have a definite
physical meaning and are formally licit.

The actual lack of synchrony between clocks at different radii can be
considered verified if one accepts the evidence from cyclotron experiments
showing particle lifetimes varying as predicted for radii associated with
different particle trajectories. There is an additional synchronization
problem, however, even for a set of clocks fixed at the same radius of a
rotating frame. Using the Einstein clock synchronization method with light
rays sent out around the disk circumference leads to a time difference (lack
of synchronization) between a clock at 0 degrees and a clock at 360 degrees 
\cite{weber}. Further, there is the well-known Sagnac effect, discovered in
1913 \cite{sagnac}, and now used as a basis for laser gyroscopes\cite
{stedman}.

The simplest explanation for this effect attributes it to the time lag
accumulated along any round trip about the central axis of the rotating disk 
\cite{arzelies}\cite{anandan}\cite{rt}.

In general what we have seen is that, in the case of a steadily rotating
disk, the only phenomenology truly irreconcilable with an old style absolute
space and time interpretation is in the behaviour of clocks. Once this is
established (and confirmed by experiment) other geometrical ''paradoxes''
and difficulties can perfectly be managed in a special relativistic frame;
one must simply be careful not to loose contact with the operative
determination of the physical quantities one is interested in.

\section{Acknowledgment}

I am indebted to my colleague Guido Rizzi for many valuable discussions and
precious suggestions.

\end{document}